\documentclass[prl,amsmath,twocolumn,superscriptaddress,wasysym]{revtex4}
\input epsf
\usepackage{epstopdf}
\usepackage{times}
\usepackage{graphicx}
\usepackage{color}
\def\beq{\begin{equation}}
\def\eeq{\end{equation}}
\def\beqn{\begin{eqnarray}}
\def\eeqn{\end{eqnarray}}

\begin{document}
\title{Dynamics after a sweep through a quantum critical point}

\author{Frank Pollmann}
\affiliation{Department of Physics, University of California, Berkeley CA 94720}

\author{ Subroto Mukerjee}
\affiliation{Department of Physics, University of California, Berkeley CA 94720}
\affiliation{Materials Science Division, Lawrence Berkeley National Laboratory, Berkeley CA 94720}
\affiliation{Department of Physics, Indian Institute of Science, Bangalore 560012, India}

\author{ Andrew G. Green}
\affiliation{School of Physics and Astronomy, University of St.~Andrews, North Haugh, St. Andrews KY16 9SS Scotland, UK}

\author{Joel E. Moore}
\affiliation{Department of Physics, University of California, Berkeley CA 94720}
\affiliation{Materials Science Division, Lawrence Berkeley National Laboratory, Berkeley CA 94720}


\date{\today}
\begin{abstract}
The coherent quantum evolution of a one-dimensional many-particle system after slowly sweeping the Hamiltonian through a critical point is studied using a generalized quantum Ising model containing both integrable and non-integrable regimes.  It is known from previous work that universal power laws of the sweep rate appear in such quantities as the mean number of excitations created by the sweep.  Several other phenomena are found that are not reflected by such averages: there are two scaling regimes of the entanglement entropy and a relaxation that is power-law in time rather than exponential.  The final state of evolution after the quench is not characterized by any effective temperature, and the Loschmidt echo converges algebraically for long times, with cusplike singularities in the integrable case that are dynamically broadened by nonintegrable perturbations.
\end{abstract}

\maketitle

A many-particle quantum system evolving at zero temperature can demonstrate various forms of approach to equilibrium even with no loss of phase coherence.  This phenomenon has been studied in most detail experimentally~\cite{Sadler:2006p312} and theoretically~\cite{Gritsev:2007p601,Lamacraft:2007p841,Lauchli:2008p527,Rossini:2009p127204,Faribault:2009p541,Hastings:2008} for systems prepared by a quantum quench across a phase transition.  A system is prepared in the ground state for certain parameter values, which are then rapidly changed to values for which the ground state is in a different phase.  Ultracold atomic systems are especially valuable for experiments in this area because they can be treated as closed quantum systems on rather long time scales compared to the basic dynamical time scales of the system.  Two basic notions in the literature are that many properties equilibrate after the quench exponentially in time and that the system thermalizes (the final state can be described by an effective temperature).  The notion of thermalization is generally not a well defined one for integrable systems and only non-integrable systems in the thermodynamic limit can be described by an effective temperature after a quench~\cite{Rigol:2008:854}.  Non-integrability alone may not be a sufficient condition for thermalization in finite systems~\cite{Biroli:2009}.

This paper concerns the coherent zero-temperature dynamics of states prepared through a different process that has also attracted recent attention~\cite{Kibble:1976p1387,Zurek:1985p505,Dziarmaga:2005p731,Polkovnikov:2005p821,Sengupta:2008p528}: rather than quenching instantly across a quantum phase transition, the Hamiltonian is changed smoothly across the transition at a constant rate $\Gamma$.  For a second-order transition (critical point), various averaged physical quantities such as the excitation density and energy show power laws in rate $\Gamma$ as $\Gamma \rightarrow 0$, with exponents determined by the quantum critical point's universal physics.  The system evolves after such a sweep to a steady state for some quantities, such as the Loschmidt echo defined below, but its energy distribution remains non-thermal.   The main result of this Letter, using a generalized quantum Ising model as an example, is that many features of the resulting evolution including the approach to equilibrium differ from the case of an instant quench and are not captured by the simple averaged quantities studied previously.  Integrable and non-integrable systems evolve differently and our results suggest a sharp dynamical probe of these differences.

\begin{figure}
\begin{center}
\includegraphics[width=75mm]{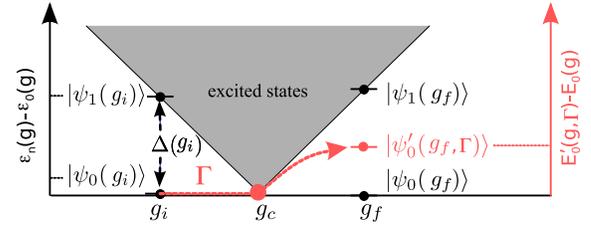}
\caption{The panel shows the energy levels $\epsilon_n(g)$ and gaps of eigenstates $|\psi_n(g)\rangle$. The energy density $E_0^{\prime}(g,\Gamma)$ of a wave function $|\psi_0^{\prime}(g,\Gamma)\rangle$ resulting from a linear sweep at a constant rate $\Gamma$ and the ground state energy density  $E_0(g)$ are shown in red.}
\label{FIG:gap}
\end{center}
\end{figure}

We consider a system with a tunable parameter $g$ that becomes critical at a value $g_c$
and study the dynamics as the Hamiltonian is swept from one side of the critical point to the other. A particular example of such a system is the generalized quantum Ising model in a field described by the Hamiltonian
\begin{equation}
H=J\sum_{i=1}^N\left[\sigma_i^x-\sigma_i^z\sigma_{i+1}^z +g\left(\cos\phi\sigma_i^x+\sin\phi\sigma_i^z\right)\right].
\label{eq:H}
\end{equation}
The system is critical at $g=g_c=0$ for any value of $\phi$. For $\phi=0$, the model Hamiltonian is the Ising model in a purely transverse field (the ``quantum Ising model'') which is integrable. For any $\phi\ne0$, the lattice model is no longer integrable (except at the critical point where $g=0$).
 We fix the value of $\phi$ and do not write the $\phi$ dependence of the quantities explicitly in the discussion below.
The correlation length goes as $\xi(g) \sim |g-g_c|^{-\nu}$ and the gap as $\Delta(g) \sim \xi^{-z} \sim |g-g_c|^{z\nu}$, close to the critical point, where $z$ is the dynamical exponent and $\nu$ the correlation length exponent (there are two values of $\nu$ depending on $\phi$, $\nu=1$ for $\phi=0,\pi$ and $\nu=8/15$ otherwise).  At each $g$, the system has a ground state $|\psi_0(g)\rangle$ with energy $\epsilon_0(g)$ and excited states $|\psi_n(g) \rangle$; the lowest excited state $|\psi_1(g) \rangle$ has energy $\Delta(g)$ (FIG.~\ref{FIG:gap}).

We study the dynamics of energy, entanglement entropy, and wavefunction overlap during and after an adiabatic sweep across a quantum critical point. We derive several analytic expressions and compare to numerical results for a sweep through the critical point of the Hamiltonian (\ref{eq:H}).
We assume that the parameter $g$ depends on time $t$ as
$g_t = g_i - \Gamma t/J$,
where $g_i$ is some initial value of $g$ and $\hbar = 1$, until some final value $g_f$ is reached, after which the Hamiltonian is constant for some ``wait period'' . We are interested in the adiabatic limit $\Gamma \rightarrow 0$. 
For the numerical calculations, we use the recently introduced \emph{infinite Time-Evolving Block Decimation} (iTEBD) algorithm \cite{Vidal:2007p070201}.
This method uses variational wavefunctions based on matrix product states (MPS) and exploits translational invariance for efficient simulation of infinite 
1D systems.
Its errors result from finite entanglement rather than finite size~\cite{Tagliacozzo:2008p024410,Pollmann:2009p255701}.  Away from a critical point, the entanglement entropy of the \emph{exact} wave function is finite, and the iTEBD algorithm becomes very accurate.

\begin{figure}
\begin{center}
\includegraphics[width=65mm]{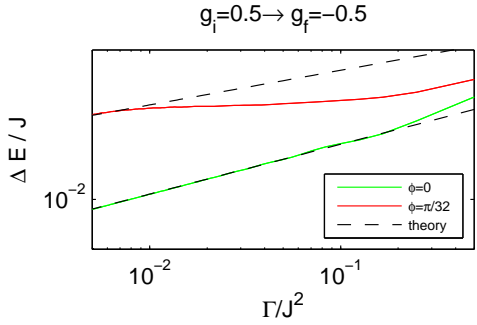}
\caption{The difference $\Delta E$ between the energy density $E_0'(g_f,\Gamma)$ of the final state after sweeping with a rate $\Gamma$ through the critical point and the ground state energy density$E_0(g_f)$. The dashed lines show the expected asymptotic  behaviors $\Delta E\sim\sqrt{\Gamma}$ ($\phi=0$) and $\Delta E \sim \Gamma^{8/23}$ ($\phi > 0$) resulting from different correlation length exponents in the ``thermal'' and ``magnetic'' directions.}
\label{FIG:E}
\end{center}
\end{figure}

The expected energy density $E_0'(g_t,\Gamma)$ of $|\psi'_0(g_t,\Gamma)\rangle$ at a time $t$ (i.e., the expectation value of the Hamiltonian with coupling $g_t$ in the time-evolved state) when the system has crossed the critical point is a natural quantity to consider. 
We use a scaling relation for slow sweeps in this model derived in Refs.~\cite{Polkovnikov:2005p821,Dziarmaga:2005p731}.  
Let the system be in the state $|\psi_0(g_i)\rangle$ at $t=0$ and change $g_t$ linearly with rate $\Gamma \rightarrow 0$.
Whilst $g_t$ is on the same side of the critical point as $g_i$, the system will be in state $|\psi_0(g_t)\rangle$ with exponential accuracy as guaranteed by the adiabatic theorem. However, 
adiabaticity breaks down
at the critical point since the gap vanishes, and once $g_t$ is on the other side, the system is in a state $|\psi'_0(g_t,\Gamma)\rangle$. Thus, excitations (well-defined in the $\phi=0$ model) are created in the system upon crossing the critical point. The number of excitations is $n_{\text{ex}}=C\Gamma^{d\nu/(z\nu+1)}$ \cite{Polkovnikov:2005p821},
where $d$ is the spatial dimension. 
Note that the scaling parameter in the above equation is the rate $\Gamma$ and not $|g-g_c|$ since $n_{\text{ex}}$ is non-zero only as a consequence of sweeping across the critical point at a finite rate. Since the excitations in the transverse Ising model can be interpreted as free domain wall excitations,  the energy density of the final state is expected to be proportional to $n_{\text{ex}}$, so
\begin{equation}
\Delta E =E_0'(g_t,\Gamma)- E_0(g_t) \sim n_{\text{ex}} \Delta(g_t)\sim \Gamma^{d\nu/(z\nu+1)} \Delta(g_t).
\label{Eq:scalener}
\end{equation}
The gap $\Delta(g_t)$ depends only on the instantaneous value of $g_t$ and does not scale with $\Gamma$.  Eqn.~(\ref{Eq:scalener}) follows from noting that as $\Gamma \rightarrow 0$, there are a small number of excitations that exist only in a small band of states of vanishing width above $|\psi_1(g) \rangle$.
Numerical results are shown in FIG.~\ref{FIG:E}. The energy difference $\Delta E$ between the actual ground state and the state after the sweep for the transverse Ising model ($\phi=0$) is in a good approximation proportional to $\sqrt{\Gamma}$. The critical exponents for the transverse Ising model are $z=1$ and $\nu=1$ and $n_{\text{ex}}\propto\sqrt{\Gamma}$ (see also Refs.~\cite{Dziarmaga:2005p731,Zurek:1985p505}). Thus the numerical results are consistent with the scaling in Eqn.~(\ref{Eq:scalener}).

\begin{figure}
\begin{center}
\includegraphics[width=65mm]{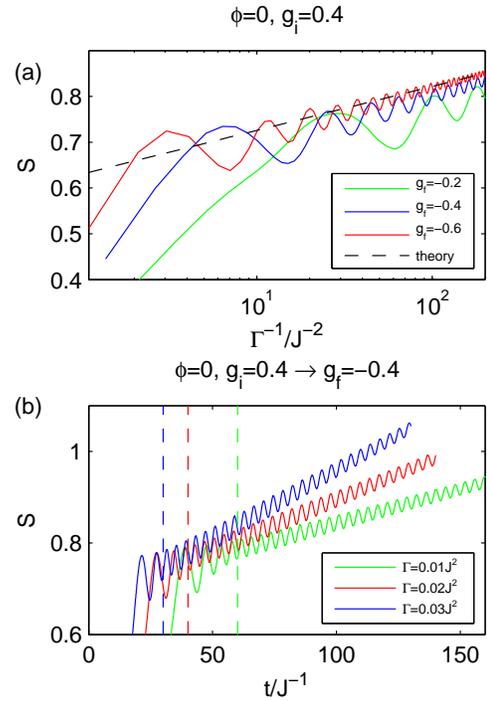}
\caption{(a) The entanglement entropy as a function of the inverse rate.  The dashed line in the upper panel shows the expected asymptotic behavior $S=-1/24\ln(\Gamma)+\text{const}$.  (b) Entanglement  entropy as a function of time. The dashed lines indicate the time at which the final value of ($g_f=-0.4$) has been reached, and the Hamiltonian remains unchanged thereafter.}
\label{FIG:S}
\end{center}
\end{figure}

We now consider the entanglement entropy between the left and right halves of the infinite system: the Hilbert space is partitioned so that all sites to the left of some bond are in one subsystem and all sites to the right are in the other.   We find one scaling law immediately after the sweep and another in the time dependence of the swept state under the final Hamiltonian, in addition to oscillatory behavior.  For critical points with conformal invariance ($z=1$) in one dimension ($d=1$), the entanglement entropy of a ground state with large finite correlation length diverges as $S = \frac{c}{6} \log \xi + \ldots$,
\label{Eq:entscal}
where $c$ is the central charge of the critical point \cite{Calabrese:2004pP06002}. To obtain the entanglement entropy, we have to find the correlation length of the state $|\psi'_0(g_t,\Gamma)\rangle$. The process of sweeping generates a gap in the system $\Delta'(g_t,\Gamma)$ different from $\Delta(g_t)$ just as $|\psi'_0(g_t,\Gamma)\rangle$ is different from $|\psi_0(g_t)\rangle$. Polkovnikov has calculated the scaling form of the ``typical gap'' $\Delta'(g_t,\Gamma)$, which is $
\Delta' \sim \Gamma^{z\nu/(z\nu+1)}.$
Now, combining this with the correlation length $\xi' \sim \Delta'^{(-z)}$ and using $z=1$,
we obtain
\begin{equation}
S = -\frac{c\nu}{6(\nu+1)} \log \Gamma+\text{const}.
\label{Eq:scalent}
\end{equation}
For the transverse Ising model with central charge $c=1/2$, we find that $S=-1/24\ln(\Gamma)+\text{const}$, which is consistent with the overall slope of the numerically calculated entanglement entropy for the  transverse Ising model ($\phi=0$) just after the sweep in FIG.~\ref{FIG:S}(a).  A similar expression for the transverse Ising model can also be obtained for the entanglement entropy of a finite block embedded in an infinite chain~\cite{Cincio:2007p052321}.


If the swept state continues to evolve in the (now constant) final Hamiltonian, the entanglement entropy oscillates around a linearly increasing mean (FIG.~\ref{FIG:S}(b)); this linear increase is as predicted by Calabrese and Cardy for a ``global quench''~\cite{Calabrese:2007p10004}, but in the swept case the slope depends on the number of excitations created during the sweep.
In our numerical study of the weakly nonintegrable model with $\phi = \pi/32$, entanglement initially grows linearly with time as in the integrable case but then, after a time related to the interactions between ``excitations'' of the integrable model, rapidly increases and makes convergence much more difficult.

As a consequence of the fact that state $|\psi'_0(g_t,\Gamma)\rangle$ includes excited states, both $S$ and the expectation values of operators that do not commute with the Hamiltonian will oscillate as a function of the sweep rate and time.  While the system is on the same side of the critical point as $g_i$, the state will evolve adiabatically.
Once the critical point is reached (at time $t_c$), the expectation value of any observable that mixes the ground state with the band of excited states oscillates with phase
${d\theta(t) \over dt} \approx \Delta(g_t).$
Using the scaling of the gap 
$\Delta(g_t) \sim |g_t-g_c|^{z\nu} = \Gamma^{z\nu}(t-t_c)^{z\nu}$,
and integrating with respect to time,  we obtain
\begin{equation}
\theta(t)\sim \Gamma^{z\nu}(t-t_c)^{z\nu+1} \sim \frac{\Delta(g_t)^{(z\nu+1)/z\nu}}{\Gamma}.
\label{Eq:scalosc}
\end{equation}
These findings are consistent with the numerical simulations of the sweep in the transverse Ising model in FIG.~\ref{FIG:S}. The entanglement entropy $S$ shows oscillations in real time (FIG.~\ref{FIG:S}(b)) which have a constant frequency equal to the gap $\Delta(g)$. The oscillations in the rate dependence in  FIG.~\ref{FIG:S}(a) are consistent with Eqn.~(\ref{Eq:scalosc}).

\begin{figure}
\begin{center}
\includegraphics[width=65mm]{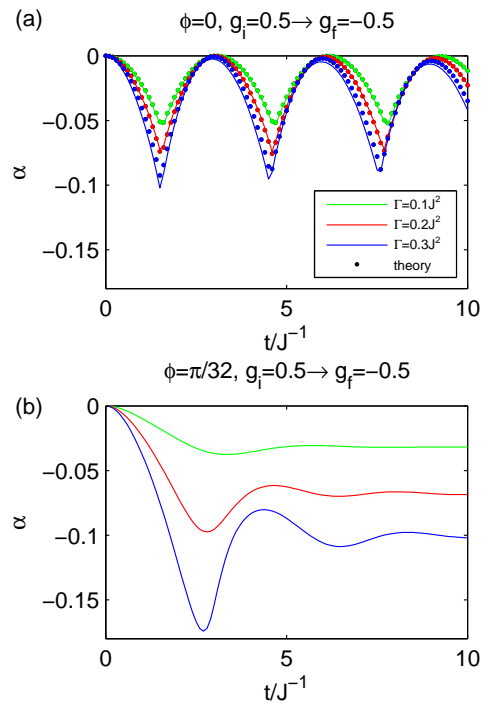}
\caption{Exponent $\alpha$ in the overlap  $\left| \langle \psi'_0(g_f,\Gamma) | \psi'_0(g_f,\Gamma,t) \rangle \right|^2=\exp(\alpha L)$. The overlap is taken of  the wavefunction $ | \psi'_0(g_f,\Gamma) \rangle$ (immediately after the ramping from $g=g_i$ to $g=g_f$ with the rate parameter $\Gamma$) and the wavefunction $ | \psi'_0(g_f,\Gamma,t) \rangle$ (after an additional evolution for the time $t$ at fixed $g=g_f$). The dots in the upper ($\phi=0$) panel show the analytical results using Eqn.~(\ref{alphaint}).}
\label{FIG:O}
\end{center}
\end{figure}

In order to study these oscillations and their damping without reference to a particular observable, we use a version of the Loschmidt echo (see Ref.~\onlinecite{Venuti:2009p09070683} and references therein).  We square the inner product between the wavefunction immediately after the sweep and the wavefunction after an additional wait period of time $t$:
\beq
\left| \langle \psi'_0(g_f,\Gamma) | \psi'_0(g_f,\Gamma,t) \rangle \right|^2 = |\langle e^{-i H_f t} \rangle_{\psi'_0(g_f,\Gamma)} |^2 = e^{- \alpha(t) L}.\label{eq:over}
\eeq
Here we have defined $\alpha(t)$ because this overlap typically goes to zero exponentially for translation-invariant states in the thermodynamic limit (i.e., when the length of the chain $L\rightarrow\infty$) unless $ | \psi'_0(g_f,\Gamma) \rangle$ and $| \psi'_0(g_f,\Gamma,t) \rangle$ are identical.  This overlap gives a direct probe of the magnitude and decay of quantum oscillations in the many-body state and it is suitable for computation using iTEBD~\cite{Vidal:2007p070201}.  It has been shown to determine the statistics of work done in a quantum quench~\cite{Silva:2009p120603}.  Our main results are that the overlap oscillations have an unusual cusp structure, arising from the integrability of the transverse Ising model, and decay even in the exact zero-temperature evolution of an infinite system (because there is a continuum of excitation energies) algebraically rather than exponentially.

An analytical approach can be developed from independent Landau-Zener tunneling~\cite{Landau:1958, Zener:1932p696} at each $k$~\cite{Dziarmaga:2005p731}.  In terms of the Bogoliubov excitations (see Ref.~\onlinecite{Dziarmaga:2005p731} for details), the wavefunction $|\psi'_0(g_f,\Gamma)\rangle$ for the transverse Ising model ($\phi=0$) is a product $|\psi'_0(g_f,\Gamma)\rangle = \prod_k (u_k |0\rangle + v_k |k,-k\rangle)$,
where $|0\rangle$ is the vacuum and the state $|k,-k\rangle$ contains a pair of quasiparticles with pseudomomenta $(k,-k)$.
The Landau-Zener formula gives, for the small $k$'s that dominate the tunneling for slow rate,
\beq
|v_k|^2 = P_k = 1 - |u_k|^2 = \exp\left(-{2 \pi J^2 k^2 \over \Gamma}\right).
\label{landauzener}
\eeq
During the wait period, each fixed-k wavefunction $|k,-k\rangle$ has an energy $\Delta_f(k) = 2 \sqrt{\Delta^2+4 J^2 gk^2}$ (this is an approximation for low energies), with lattice spacing $a = 1$.  So up to an overall phase
\beq
| \psi'_0(g_f,\Gamma,t) \rangle = \prod_k (u_k |0\rangle + e^{-i\Delta_f(k) t} v_k |k,-k \rangle)
\eeq
Now, using Eqn.~(\ref{landauzener}), the squared overlap in Eqn.~(\ref{eq:over}) is rewritten as 
\beq
\prod_k  \left[1+4\sin^2\left(\frac{\Delta_f(k) t}2\right)P_k(P_k-1)\right].
\eeq
In the continuum limit, the logarithm of this product becomes an integral for $\alpha(t)$.  Taking the momentum cutoff to $\infty$,
\beq
\alpha(t)= {1 \over 2 \pi} \int_0^\infty\,dk\,\log\left[1+4\sin^2\left(\frac{\Delta_f(k) t}2\right)P_k(P_k-1)\right].
\label{alphaint}
\eeq
This 
is compared to the numerical results in FIG.~\ref{FIG:O}(a).  The cusplike minima arise from the momentum value $k^*$ where the tunneling probability $P_{k^*} = 1/2$.  These give a singularity in $\alpha(t)$ at times $\Delta_f(k^*) t = (2n + 1) \pi$.
This singular behavior is smeared out in the nonintegrable model at $\phi = \pi/32$, which does not have sharp excitations at finite energy (FIG.~\ref{FIG:O}(b)).  This gives a probe of integrable versus nonintegrable behavior that could be useful for experiments~\cite{Kinoshita:2006p900}.  Asymptotics of Eqn.~(\ref{alphaint}) lead to the following results:
if we have $\Gamma N \ll \Delta^2$
for some number of periods $N \geq 1$, then the overlap has peaks shifted from the points $\Delta t = \pi N$, with maxima
$\alpha_N=- {\sqrt{2} \over \pi} {g^2\Gamma^{5/2} \over J \Delta^4} \left[ {7 \over 32} - {\sqrt{2}\over 16}\right]N^2.$
At long times, $\alpha$ approaches a constant independent of the final gap: $\alpha(\infty) \approx -0.0564\sqrt{\Gamma}/J.$

In the integrable case, the amplitude of oscillations in the Loschmidt echo falls off as $1/t$ for large times, which is different from the case of a quasithermal initial state whose excitation probabilities are given by Boltzmann weights.  The latter has no cusps and damping as $1/\sqrt{t}$.  This power-law decay of the oscillations, rather than an exponential decay, is an important difference between the evolution found here and the decay of some observables after a sudden quench~\cite{Rossini:2009p127204}.  We note that the (pure) state as $t \rightarrow \infty$ does not resemble a (mixed) thermal state at any effective temperature of the final Hamiltonian, since the excited state occupation probabilities remain determined by their tunneling probability $P_k$, while their energy is $\Delta_f(k)$; the final state has more occupation of low-$k$ excited states than a thermal state.  Sweeping can also induce {\it spatial} oscillations in correlation functions of local operators~\cite{Cherng:2006p043614}.

These results are modified in several ways for  non-integrable systems.  In any system, the Loschmidt echo depends on both the initial state and the Fourier transform of the energy spectrum.  The latter is known to change between integrable and non-integrable systems:
one sharp difference in spectral properties is revealed  through the smearing out of cusps noted above (FIG.~\ref{FIG:O}(b)).  More generally, the measurement of a Loschmidt echo could be used to probe the difference in energy level statistics between integrable (Poisson) and non-integrable (Wigner-Dyson) systems.

Local observables are also likely to behave differently in integrable and non-integrable systems, e.g., decay of {\it local} observables to a thermal distribution is expected in the non-integrable case.  The dynamical crossover to non-integrable behavior was observed here as a rapid increase of entanglement entropy and the destruction of cusps in the Loschmidt echo.  Accessing the long-time regime and approach to equilibrium in the non-integrable case is a major challenge for numerical methods.

The authors acknowledge conversations with A.~Polkovnikov and H.~Saleur and support from DARPA OLE (F.~P.), DOE (S.~M.), the Miller Institute and the Royal Society (A.~G. ~G.), and NSF DMR-0804413 (J.~E.~M.),

\end{document}